\documentclass[aip,pof,reprint,unsortedaddress,amsmath,amssymb]{revtex4-1}

\usepackage{graphics,graphicx}% Include figure files
\usepackage{dcolumn}% Align table columns on decimal point
\begin{document}

\title{Two-point velocity average of turbulence: statistics and their implications}

\author{Hideaki Mouri}
\email{hmouri@mri-jma.go.jp}
%\affiliation{Meteorological Research Institute, Nagamine, Tsukuba 305-0052, Japan}

\author{Akihiro Hori}
\altaffiliation[Also at ]{Meteorological and Environmental Sensing Technology, Inc., Nanpeidai, Ami 300-0312, Japan.}
\affiliation{Meteorological Research Institute, Nagamine, Tsukuba 305-0052, Japan}

%%%%%%%%%%%%%%%%%%%%%%%%%%%%%%%%%%%%%%%%%%%%%%%%%%%%%%%%%%%%%%%%%%%%%%%%%%%%%%%%%%%
%                                                                                 %
%  To the publisher:                                                              %
%                                                                                 %
%  You may find the expression ``$10^1$'' etc.                                    %
%  Please do not replace it with, e.g., ``10''.                                   %
%                                                                                 %
%                                                                                 %
%  If you want to modify any of our figures (letter size, letter position, ...),  %
%  please let us know.                                                            %
%  We would modify the figure by ourselves,                                       %
%  in order to avoid possible problems occuring during the modification.          %
%                                                                                 %
%  The widths of the figures should be as large as possible,                      % 
%      so far as they are single-columns figures.                                 %
%  The widths of the figures are all the same (say, 7cm).                         %    
%                                                                                 %
%%%%%%%%%%%%%%%%%%%%%%%%%%%%%%%%%%%%%%%%%%%%%%%%%%%%%%%%%%%%%%%%%%%%%%%%%%%%%%%%%%%

\date{3 October 2010}

\begin{abstract}
For turbulence, although the two-point velocity difference $u(x+r)-u(x)$ at each scale $r$ has been studied in detail, the velocity average $[u(x+r)+u(x)]/2$ has not thus far. Theoretically or experimentally, we find interesting features of the velocity average. It satisfies an exact scale-by-scale energy budget equation. The flatness factor varies with the scale $r$ in a universal manner. These features are not consistent with the existing assumption that the velocity average is independent of $r$ and represents energy-containing large-scale motions alone. We accordingly propose that it represents motions over scales $\ge r$ as long as the velocity difference represents motions at the scale $r$.
\end{abstract}

\maketitle

\section{INTRODUCTION} \label{s1}

Suppose that a velocity component $u(x,t)$ is obtained at time $t$ along a one-dimensional cut $x$ of a turbulent flow. The ensemble average $\langle u \rangle$ has been subtracted, so as to have $\langle u \rangle = 0$ anywhere below. Then, the two-point velocity difference $u_-$ and average $u_+$ are defined at each scale $r$ as 
%_______________________________________________________________________________________
\begin{subequations}
\label{eq1}
\begin{align}
& u_-(r,x,t) = u(x+r,t)-u(x,t),
\\
& u_+(r,x,t) = \frac{u(x+r,t)+u(x,t)}{2}.
\end{align}
\end{subequations}
%_______________________________________________________________________________________
While $u_-$ has been studied in detail,\cite{k41a,sa97} $u_+$ has not thus far. We are interested in $u_+$ and are to study its statistical features.

Sreenivasan and Dhruva\cite{sd98} studied the probability density distribution of $u_+(r)$ in atmospheric turbulence. The shape of the distribution was close to that of $u$, regardless of the scale $r$.

Tatsumi and Yoshimura\cite{ty04,ty07} studied the probability density distribution of $u_+(r)$ at each scale $r$ in a theoretical manner, by applying a closure approximation to homogeneous isotropic turbulence.

Hosokawa\cite{ho07} found an exact relation for turbulence that is homogeneous in the $x$ direction. Since $\langle u(x+r)^3 \rangle = \langle u(x)^3 \rangle$, where $\langle \cdot \rangle$ denotes an ensemble average, Eq.~(\ref{eq1}) yields
%_______________________________________________________________________________________
\begin{equation}
\label{eq2}
\langle u_-^3(r) \rangle = -12 \langle u_+^2(r) u_-(r) \rangle.
\end{equation} 
%_______________________________________________________________________________________
Through this equation, Kolmogorov's\cite{k41b} four-fifth law for the longitudinal velocity at small scale $r$ in the inertial range
%_______________________________________________________________________________________
\begin{subequations}
\begin{equation}
\label{eq3a}
\langle u_-^3(r) \rangle = -\frac{4}{5} \langle \varepsilon \rangle r
\end{equation}
%_______________________________________________________________________________________
is equivalent to
%_______________________________________________________________________________________
\begin{equation}
\label{eq3b}
\langle u_+^2(r) u_-(r) \rangle = \frac{1}{15} \langle \varepsilon \rangle r.
\end{equation}
\end{subequations}
%_______________________________________________________________________________________
Here $\varepsilon$ is the energy dissipation rate. Kholmyansky and Tsinober\cite{kt08} confirmed Eq.~(\ref{eq3b}) in atmospheric turbulence. Exact relations were also found by Hill\cite{hi02} and Germano.\cite{ge07}

These studies tend to assume that, although $u_-(r)$ represents motions at the scale $r$,\cite{k41a,sa97} $u_+(r)$ does not depend on $r$ and represents energy-containing large-scale motions alone.\cite{sd98,ho07,kt08,hi02} If this were the case, $\langle u_+^2(r) u_-(r) \rangle \ne 0$ for $r$ in the inertial range [Eq.~(\ref{eq3b})] would imply a correlation between small-scale motions in the inertial range and large-scale motions in the energy-containing range.\cite{ho07,kt08} Such a correlation is of interest because still controversial is whether or not small-scale motions are statistically independent of large-scale motions.\cite{sd98,m06,ll59,o62,p93}

However, it is necessary to reconsider the assumption for $u_+(r)$. We find that $u_+^2$ satisfies an exact scale-by-scale energy budget equation that involves Eq.~(\ref{eq3b}). We also use experimental data to find that $\langle u_+^4(r) \rangle / \langle u_+^2(r) \rangle ^2$ varies with $r$ in a universal manner. Thus, albeit dominated by large-scale motions, $u_+(r)$ reflects motions at the scale $r$.\cite{ty04,ty07} We accordingly propose an assumption that $u_+(r)$ represents motions over scales $\ge r$. Since the correlation of $u_+^2(r)$ with $u_-(r)$ in Eq.~(\ref{eq3b}) is attributable to motions at the scale $r$, there is no more need to invoke a correlation between small- and large-scale motions. We discuss this and other implications.

%_______________________________________________________________________________________
\begingroup
\squeezetable
\begin{table}[htbp]
\caption{\label{t1} {\footnotesize
Parameters of grid turbulence (GT), boundary layer (BL), and jet (J) taken from Ref. \onlinecite{mht09}: kinematic viscosity $\nu$, mean rate of energy dissipation $\langle \varepsilon \rangle$, Kolmogorov velocity $u_K$, rms velocity fluctuations $\langle u^2 \rangle^{1/2}$ and $\langle v^2 \rangle^{1/2}$, skewness factor $\langle u^3 \rangle / \langle u^2 \rangle^{3/2}$, flatness factors $\langle u^4 \rangle / \langle u^2 \rangle^2$ and $\langle v^4 \rangle / \langle v^2 \rangle^2$, Kolmogorov length $\eta$, correlation lengths $L_u$ and $L_v$, and Reynolds number Re$_{\lambda}$ for the Taylor microscale $\lambda = [2 \langle v^2 \rangle / \langle (\partial _x v)^2 \rangle ]^{1/2}$.}}

\begin{ruledtabular}
\begin{tabular}{lcccc}
\noalign{\smallskip}
Quantity                                                                       & Units           & GT     & BL     & J      \\ 
\hline
\noalign{\smallskip}
$\nu$                                                                          & cm$^2$ s$^{-1}$ &$0.143$ &$0.143$ &$0.139$ \\
$\langle \varepsilon \rangle = 15 \nu \langle (\partial _x v)^2 \rangle /2$    & m$^2$ s$^{-3}$  &$2.81$  &$12.6$  &$2.60$  \\
$u_K = (\nu \langle \varepsilon \rangle)^{1/4}$                                & m s$^{-1}$      &$0.0796$&$0.116$ &$0.0776$\\
$\langle u^2 \rangle^{1/2}$                                                    & m s$^{-1}$      &$0.696$ &$2.37$  &$1.56$  \\
$\langle v^2 \rangle^{1/2}$                                                    & m s$^{-1}$      &$0.683$ &$1.96$  &$1.36$  \\
$\langle u^3 \rangle / \langle u^2 \rangle^{3/2}$                              &                 &$+0.08$ &$-0.10$ &$-0.04$ \\
$\langle u^4 \rangle / \langle u^2 \rangle^2$                                  &                 &$3.00$  &$2.69$  &$2.60$  \\
$\langle v^4 \rangle / \langle v^2 \rangle^2$                                  &                 &$2.98$  &$3.05$  &$3.05$  \\
$\eta = (\nu ^3 / \langle \varepsilon \rangle )^{1/4}$                         & cm              &$0.0180$&$0.0123$&$0.0179$\\
$L_u  = \int^{\infty}_{0} \langle u(x+r)u(x) \rangle dr / \langle u^2 \rangle$ & cm              &$17.5$  &$43.0$  &$128.$  \\
$L_v = \int ^{\infty}_{0} \langle v(x+r)v(x) \rangle dr / \langle v^2 \rangle$ & cm              &$4.46$  &$5.68$  &$10.2$  \\
Re$_{\lambda}  = \langle v^2 \rangle^{1/2} \lambda / \nu$                      &                 &$285$   &$1103$  &$1183$  \\

\end{tabular}
\end{ruledtabular}
\end{table}
\endgroup
%_______________________________________________________________________________________

\section{EXPERIMENTAL DATA} \label{s2}

Experimental data of grid turbulence, boundary layer, and jet are used here. They were obtained with a wind tunnel in our recent work.\cite{mht09} We outline the experiments and the data. Their parameters are shown in Table \ref{t1}.

\subsection{Grid turbulence} \label{s2a}

For the grid turbulence, we set a grid across the wind tunnel. The grid was made of rods with $0.04 \times 0.04$\,m$^2$ in cross section and $0.20$\,m in separation. At $x_{\rm wt} = 3.5$\,m downstream of the grid, a hot-wire anemometer was used to measure the streamwise velocity $U+u(t_{\rm wt})$ and the spanwise velocity $v(t_{\rm wt})$. Here $U$ is the average, whereas $u(t_{\rm wt})$ and $v(t_{\rm wt})$ are temporal fluctuations. The data are long so that their statistical significance is high. Also at $x_{\rm wt} = 3.25$\,m and $3.75$\,m, we measured $U+u(t_{\rm wt})$. This measurement was not simultaneous with the above measurement but was under the same condition.

The temporal fluctuations $u(t_{\rm wt})$ and $v(t_{\rm wt})$ measured at position $x_{\rm wt}$ are converted into the spatial fluctuations $u(x)$ and $v(x)$ at time $t$, by using Taylor's frozen-eddy hypothesis
%_______________________________________________________________________________________
\begin{equation}
\label{eq4}
x = -U t_{\rm wt} \quad \mbox{and} \quad t = \frac{x_{\rm wt}}{U}.
\end{equation}
%_______________________________________________________________________________________
Since the grid turbulence was stationary in $t_{\rm wt}$ and was decaying in $x_{\rm wt}$, it is homogeneous in $x$ and is decaying in $t$.\cite{m06,note0} The data at $x_{\rm wt} = 3.25$\,m and $3.75$\,m are used to estimate $\partial _t \langle u^2 \rangle$ and so on at the time $t$ corresponding to $x_{\rm wt} = 3.5$\,m.

\subsection{Boundary layer and jet}

The data of the boundary layer and jet were obtained in the same manner as for the grid turbulence. For the boundary layer, roughness was set over the floor of the wind tunnel. The measurement was done in the log-law region at a streamwise position $x_{\rm wt}$ where the boundary layer had been well developed. For the jet, a nozzle was set within the wind tunnel. The measurement was done where the flow had become turbulent. Since these measurements were under stationary conditions, we use Taylor's hypothesis to obtain spatial fluctuations that are homogeneous in the $x$ direction [Eq.~(\ref{eq4})].

To explore common features, we compare the boundary layer and jet with the grid turbulence. The comparison is based on the correlation length $L_u$ (Table \ref{t1}), which represents large scales and also serves as the typical size of energy-containing eddies.

%_______________________________________________________________________________________
\begin{figure}[hbp]
\resizebox{8cm}{!}{\includegraphics*[4.5cm,9.7cm][16.5cm,26.cm]{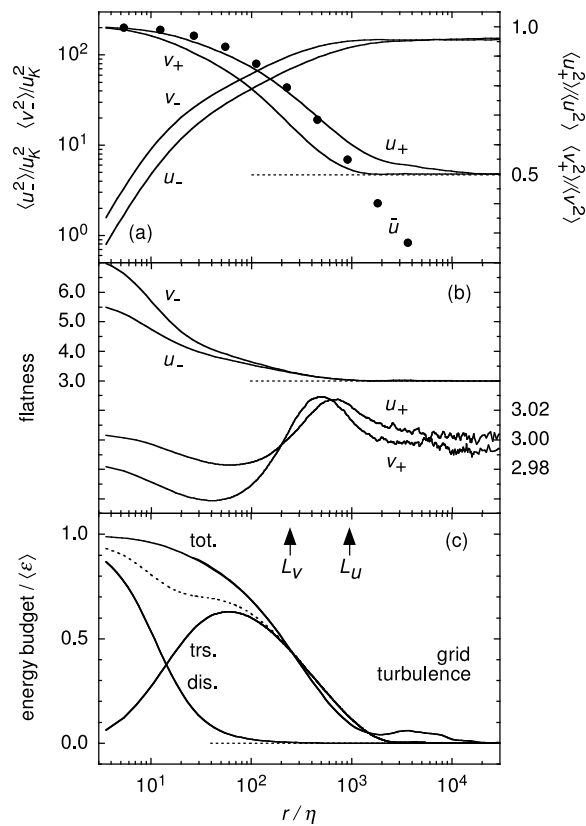}}
\caption{\label{f1} {\footnotesize Statistics for grid turbulence as a function of $r/\eta$. (a) $\langle u_+^2 \rangle / \langle u^2 \rangle$, $\langle v_+^2 \rangle / \langle v^2 \rangle$, $\langle u_-^2 \rangle/u_K^2$, and $\langle v_-^2 \rangle/u_K^2$. The circles denote $\langle \bar{u}^2 \rangle / \langle u^2 \rangle$. (b) Flatness factors $\langle u_+^4 \rangle / \langle u_+^2 \rangle^2$, $\langle v_+^4 \rangle / \langle v_+^2 \rangle^2$, $\langle u_-^4 \rangle / \langle u_-^2 \rangle^2$, and $\langle v_-^4 \rangle / \langle v_-^2 \rangle^2$. (c) Energy budget in Eq.~(\ref{eq10}). The left-hand side (``tot.'') and the second term in the right-hand side (``trs.'') are normalized by $-3 \partial_t \langle u^2 \rangle /2$. The first term (``dis.'') is normalized by $15 \nu \langle (\partial_x u)^2 \rangle$. The dotted curve denotes the sum of the terms in the right-hand side. The arrows indicate $r = L_u$ and $L_v$.}}
\end{figure} 
%_______________________________________________________________________________________
\begin{figure}[htbp]
\resizebox{8cm}{!}{\includegraphics*[4.5cm,9.7cm][16.5cm,26.cm]{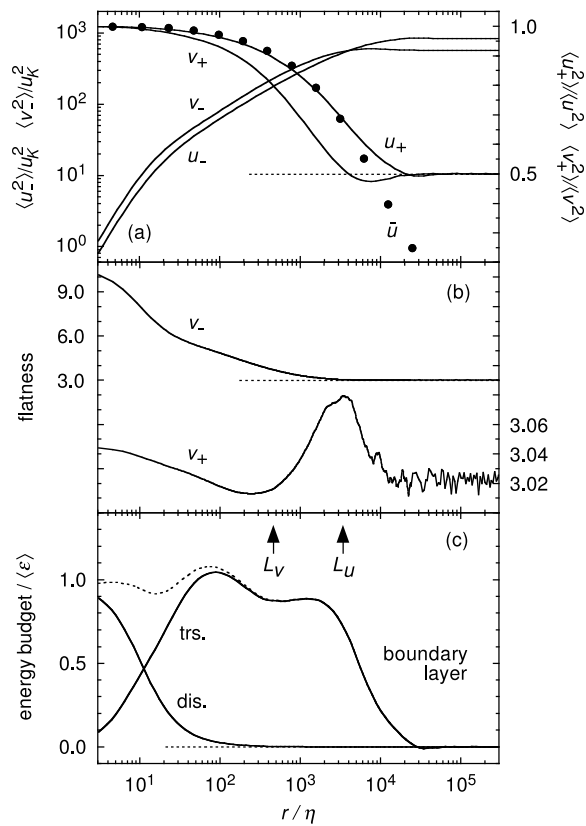}}
\caption{\label{f2} {\footnotesize Statistics for boundary layer as a function of $r/\eta$. (a) $\langle u_+^2 \rangle / \langle u^2 \rangle$, $\langle v_+^2 \rangle / \langle v^2 \rangle$, $\langle u_-^2 \rangle/u_K^2$, and $\langle v_-^2 \rangle/u_K^2$. The circles denote $\langle \bar{u}^2 \rangle / \langle u^2 \rangle$. (b) Flatness factors $\langle v_+^4 \rangle / \langle v_+^2 \rangle^2$ and $\langle v_-^4 \rangle / \langle v_-^2 \rangle^2$. (c) Energy budget in Eq.~(\ref{eq11}). The first term in the right-hand side (``dis.'') is normalized by $15 \nu \langle (\partial_x u)^2 \rangle$, and the second term (``trs.'') by $15 \nu \langle (\partial_x v)^2 \rangle /2$. The dotted curve denotes their sum. The arrows indicate $r = L_u$ and $L_v$.}}
\end{figure} 
%_______________________________________________________________________________________
\begin{figure}[hbp]
\resizebox{8cm}{!}{\includegraphics*[4.5cm,9.7cm][16.5cm,26.cm]{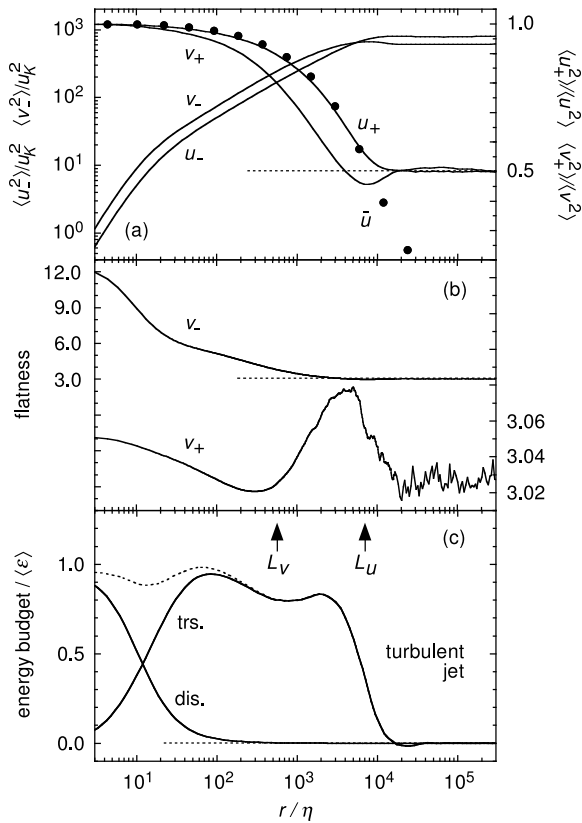}}
\caption{\label{f3} {\footnotesize Same as in Fig.~\ref{f2} but for jet.}}
\end{figure} 
%_______________________________________________________________________________________

\section{MEANING OF VELOCITY AVERAGE} \label{s3}

Usually, statistics of $u_-(r)$ and $v_-(r)$ are discussed by assuming that $u_-(r)$ and $v_-(r)$ represent motions at the scale $r$.\cite{k41a,sa97} To be exact, we also have to consider motions at scales $< r$. Their contributions to $u_-(r)$ and $v_-(r)$ are nevertheless negligible as long as $r$ is not too large.\cite{To76} This is because the energy distribution of a turbulent flow is a continuously and rapidly increasing function of the scale $r$ up to about the correlation length $L_u$.

Then, as long as $r$ is not too large for $u_-(r)$ and $v_-(r)$ to be safely assumed as representatives of motions at the scale $r$, we assume $u_+(r)$ and $v_+(r)$ as representatives of motions over scales $\ge r$. If this is the case, $u_+(r)$ and $v_+(r)$ reflect motions at the scale $r$, albeit dominated by motions at the larger scales, $> r$. Those at scales $< r$ also contribute to $u_+(r)$ and $v_+(r)$, but they are expected to be negligible as long as they are negligible in $u_-(r)$ and $v_-(r)$.

Our assumption is justified by using the experimental data to compare $\langle u_+^2(r) \rangle$ with $\langle \bar{u}^2(r) \rangle$. Here,  
%_______________________________________________________________________________________
\begin{equation}
\label{eq5}
\bar{u}(r,x,t) = \frac{1}{r} \int_{x}^{x+r} u(X,t)dX,
\end{equation}
%_______________________________________________________________________________________
which defines the motions over scales $\ge r$ because those at scales $< r$ have been smoothed away. The results are shown in Figs.~\ref{f1}(a), \ref{f2}(a), and \ref{f3}(a). We observe $\langle u_+^2(r) \rangle \simeq \langle \bar{u}^2(r) \rangle$ for $r \lesssim L_u$, where it is safe to assume $u_+(r)$ as a representative of motions over scales $\ge r$. This is not the case for $r \gtrsim L_u$, where motions at scales $< r$ are no longer negligible in $\langle u_+^2(r) \rangle$ as well as in $\langle u_-^2(r) \rangle$ and accordingly these two are constant.

Our assumption is self-consistent. For example, in the limit $r \rightarrow 0$, $u_+(r)$ and $v_+(r)$ reduce to $u$ and $v$ that represent motions over all the scales.

The scales of motions could be also defined with wave number $k$. We have
%_______________________________________________________________________________________
\begin{subequations}
\label{eq6}
\begin{align}
& \langle u_-^2(r) \rangle \propto \int^{\infty}_{0} [1-\cos(kr)] E_u(k) dk, \\
& \langle u_+^2(r) \rangle \propto \int^{\infty}_{0} [1+\cos(kr)] E_u(k) dk,
\end{align}
\end{subequations}
%_______________________________________________________________________________________
where $E_u(k)$ is the energy spectrum. This is consistent with our assumption because $1+\cos(kr)$ is maximal at $k = 0$ and falls to zero at $k = \pi / r$, where $1-\cos(kr)$ is maximal.\cite{note1} However, being localized in the $x$ space, $u_+$ or $u_-$ is not localized in the $k$ space. For the scales, our definition based on Eq.~(\ref{eq5}) is preferable.

Phenomenologically, $u_-(r)$ and $v_-(r)$ represent the deformation and rotation of an eddy with size $r$, respectively, if $r$ does not exceed the typical size of energy-containing eddies, $L_u$. This is just the $r$ range for $u_-(r)$ and $v_-(r)$ to represent motions at the scale $r$ [Figs.~\ref{f1}(a), \ref{f2}(a), \ref{f3}(a)]. Likewise, we assume $u_+(r)$ and $v_+(r)$ as the advection of an eddy with size $r$. The advection is due to sweeping by nearby eddies with sizes $\gtrsim r$, being consistent with the above assumption that motions over scales $\ge r$ are represented by $u_+(r)$ and $v_+(r)$.

Between $u_+$ and $u_-$, exact statistical relations exist if the turbulence is homogeneous in the $x$ direction.\cite{ho07} We have already mentioned $\langle u_-^3(r) \rangle = -12 \langle u_+^2(r) u_-(r) \rangle$ [Eq.~(\ref{eq2})]. In addition, since $\langle u(x)^2 \rangle = \langle u(x+r)^2 \rangle$, Eq.~(\ref{eq1}) yields
%_______________________________________________________________________________________
\begin{subequations}
\label{eq7}
\begin{align}
\label{eq7a}& \langle u_+^2(r) \rangle + \frac{\langle u_-^2(r) \rangle}{4} = \langle u^2 \rangle, \\
\label{eq7b}& \langle u_+(r) u_-(r) \rangle = 0.                                             
\end{align} 
\end{subequations}
%_______________________________________________________________________________________
The same relations exist between $v_+$ and $v_-$. For each $r$, the mean total energy $\langle u^2 \rangle$ is decomposed into the mean energies $\langle u_+^2(r) \rangle$ and $\langle u_-^2(r) \rangle$ [Eq.~(\ref{eq7a})]. Although $u_+(r)$ does not correlate with $u_-(r)$ [Eq.~(\ref{eq7b})], $u_+^2(r)$ could correlate with $u_-(r)$ [Eq.~(\ref{eq2})]. This correlation is through motions at the scale $r$, if we ignore motions at scales $< r$.

The relations in Eqs.~(\ref{eq2}) and (\ref{eq7}) are not restricted to the homogeneous turbulence. The same relations exist in turbulence that is isotropic throughout the scales.\cite{kt08,note2,note3} We nevertheless focus on the homogeneous turbulence, which is more common than the isotropic turbulence. For example, an experiment is usually under a stationary condition. The measured temporal fluctuations are converted by Taylor's hypothesis into spatial fluctuations that are homogeneous in the $x$ direction [Eq.~(\ref{eq4})].

\section{ENERGY BUDGET EQUATION} \label{s4}

The scale-by-scale energy budget is studied in terms of $u_+^2$. We consider turbulence that is homogeneous and isotropic throughout the scales and also is decaying freely in time, for which an exact equation is known:\cite{ll59,s68}
%_______________________________________________________________________________________
\begin{subequations}
\label{eq8}
\begin{align}
\label{eq8a}
&  \frac{\partial}{\partial t}
  \left[
  -\frac{3}{2} \langle u^2 \rangle
 +\frac{15}{4r^5} \int^r_0 \langle u_-^2(R) \rangle R^4 dR
  \right] \nonumber \\
& \qquad \qquad \qquad \qquad
  =\frac{15 \nu}{2r} \frac{\partial \langle u_-^2(r) \rangle}{\partial r}
  -\frac{5 \langle u_-^3(r) \rangle}{4r}.
\end{align}
%_______________________________________________________________________________________
By substituting Eqs.~(\ref{eq2}) and (\ref{eq7a}) into Eq.~(\ref{eq8a}) and also by rearranging the terms in the left-hand side, we obtain
%_______________________________________________________________________________________
\begin{align}
\label{eq8b}
&  \frac{\partial}{\partial t}
   \left[
   \frac{3}{2} \langle u^2 \rangle 
  -\frac{15}{r^5} \int^r_0 \langle u_+^2(R) \rangle R^4 dR 
   \right] \nonumber \\
& \qquad \qquad  
 =-\frac{30 \nu}{r} \frac{\partial \langle u_+^2(r) \rangle}{\partial r}
  +\frac{15 \langle u_+^2(r) u_-(r) \rangle}{r}.
\end{align}
\end{subequations}
%_______________________________________________________________________________________
Thus, $u_+^2$ satisfies an exact scale-by-scale equation. This would not exist if $u_+(r)$ were a representative of large-scale motions alone. It does represent motions at scales $\ge r$.

To clarify the meanings of Eqs.~(\ref{eq8a}) and (\ref{eq8b}), their left-hand sides are rewritten by using the relation
%_______________________________________________________________________________________
\begin{align}
\label{eq9}
  \frac{3}{2} \langle u^2 \rangle
  =
& \left. \frac{15}{4r^5} \int^r_0 \langle u_-^2(R) \rangle R^4 dR \right|_{r \rightarrow \infty} \nonumber \\
  =
& \left. \frac{15}{r^5} \int^r_0 \langle u_+^2(R) \rangle R^4 dR \right|_{r \rightarrow \infty},
\end{align}
%_______________________________________________________________________________________
which is from the asymptote $\langle u(x+r)u(x) \rangle \rightarrow 0$ for $r \rightarrow \infty$. The results are
%_______________________________________________________________________________________
\begin{widetext}
\begin{subequations}
\label{eq10}
\begin{equation}
\label{eq10a}
  \frac{\partial}{\partial t}
  \left[
  -\frac{15}{4r^5} \left. \int^r_0 \langle u_-^2(R) \rangle R^4 dR \right|_{r \rightarrow \infty}
  +\frac{15}{4r^5}        \int^r_0 \langle u_-^2(R) \rangle R^4 dR
  \right]                                
 = \frac{15 \nu}{2r} \frac{\partial \langle u_-^2(r) \rangle}{\partial r}
   -\frac{5 \langle u_-^3(r) \rangle}{4r} ,
\end{equation}

\begin{equation}
\label{eq10b}
   \frac{\partial}{\partial t}
   \left[
   \frac{15}{r^5} \left. \int^r_0 \langle u_+^2(R) \rangle R^4 dR \right|_{r \rightarrow \infty}
  -\frac{15}{r^5}        \int^r_0 \langle u_+^2(R) \rangle R^4 dR
   \right]
= -\frac{30 \nu}{r} \frac{\partial \langle u_+^2(r) \rangle}{\partial r}
  +\frac{15 \langle u_+^2(r) u_-(r) \rangle}{r} .
\end{equation}
\end{subequations}
\end{widetext}
%_______________________________________________________________________________________
These equations offer different descriptions of the same energy budget. By using Eqs.~(\ref{eq2}) and (\ref{eq7a}), the left-hand side of Eq.~(\ref{eq10b}) and the individual terms in its right-hand side are able to be obtained one-by-one from those of Eq.~(\ref{eq10a}). Thus, each two of these corresponding terms have exactly the same values. Since Eq.~(\ref{eq10a}) describes the budget of kinetic energy per unit time and mass in terms of $u_-^2$ (see below), it is natural to consider that Eq.~(\ref{eq10b}) describes the same energy budget in terms of $u_+^2$. Here $u_-^2$ or $u_+^2$ is not used as the energy itself but is used to describe its budget. Although $u_-^2$ and $u_+^2$ behave differently [Figs.~\ref{f1}(a), \ref{f2}(a), and \ref{f3}(a)], they are statistically dependent on each other through Eqs.~(\ref{eq2}) and (\ref{eq7a}).

For the grid turbulence, Fig.~\ref{f1}(c) shows values of the terms of Eqs.~(\ref{eq10a}) and (\ref{eq10b}).\cite{note4} Their corresponding terms are actually not distinguishable at all.

The left-hand side of Eq.~(\ref{eq10}) is just a weighted average of $\partial_t \langle u_-^2(R )\rangle$ or $\partial_t \langle u_+^2(R) \rangle$ over $R \ge r$, which describes the mean rate of the total loss of the energy over scales $\ge r$. It reduces to $-3 \partial_t \langle u^2 \rangle /2 = \langle \varepsilon \rangle$ as $r \rightarrow 0$ and to $0$ as $r \rightarrow \infty$. The first term in the right-hand side of Eq.~(\ref{eq10}) describes the mean rate of the energy dissipation into heat over scales $\ge r$. It reduces to $15 \nu \langle ( \partial_x u )^2 \rangle = \langle \varepsilon \rangle$ as $r \rightarrow 0$ and to $0$ as $r \rightarrow \infty$. These two are balanced with the second term, which describes the mean rate of the energy transfer across $r$ from the larger to the smaller scales.\cite{note5}

The transfer term for $u_+^2$ involves $u_-$ [Eq.~(\ref{eq10b})], which is phenomenologically related to the typical time scale $r/\langle u_-^2(r) \rangle^{1/2}$ for the energy transfer across the scale $r$ (Sec.~\ref{s5}).

Between each two of the corresponding terms of Eqs.~(\ref{eq10a}) and (\ref{eq10b}), there is a sign difference. This difference is associated with the fact that $u_-^2(r)$ tends to increase with $r$ while $u_+^2(r)$ tends to decrease with $r$ [Figs.~\ref{f1}(a), \ref{f2}(a), and \ref{f3}(a); see also Eq.~(\ref{eq7a})].

When the scale $r$ is small enough to lie in the inertial or dissipative range, the left-hand side of Eq.~(\ref{eq10a}) is able to be replaced by $\langle \varepsilon \rangle$:
%_______________________________________________________________________________________
\begin{subequations}
\label{eq11}
\begin{equation}
\label{eq11a}
  \langle \varepsilon \rangle 
= \frac{15 \nu}{2r} \frac{\partial \langle u_-^2(r) \rangle}{\partial r}
  -\frac{5 \langle u_-^3(r) \rangle}{4r}.
\end{equation}
%_______________________________________________________________________________________
This equation in turn applies to small scales of any turbulence that is isotropic over these small scales.\cite{k41b,ll59} If the turbulence is also homogeneous in the $x$ direction, Eq. (\ref{eq10b}) is likewise rewritten as
%_______________________________________________________________________________________
\begin{equation} 
\label{eq11b} 
  \langle \varepsilon \rangle
= -\frac{30 \nu}{r} \frac{\partial \langle u_+^2(r) \rangle}{\partial r}
  +\frac{15 \langle u_+^2(r) u_-(r) \rangle}{r}.
\end{equation}
\end{subequations}
%_______________________________________________________________________________________
For the boundary layer and jet, Figs.~\ref{f2}(c) and \ref{f3}(c) show values of the terms in the right-hand sides of Eqs.~(\ref{eq11a}) and (\ref{eq11b}).\cite{note4}

Within the inertial range, the right-hand sides of Eqs. (\ref{eq11a}) and (\ref{eq11b}) are dominated by the second terms. They correspond to Eqs.~(\ref{eq3a}) and (\ref{eq3b}). That is, while the transfer term for $u_-^2$ in Eq.~(\ref{eq11a}) corresponds to Kolmogorov's\cite{k41b} four-fifth law (\ref{eq3a}), the transfer term for $u_+^2$ in Eq.~(\ref{eq11b}) corresponds to Hosokawa's\cite{ho07} relation (\ref{eq3b}).

%_______________________________________________________________________________________
\begin{figure}[bp]
\resizebox{8cm}{!}{\includegraphics*[4.cm,14.7cm][16.cm,26.3cm]{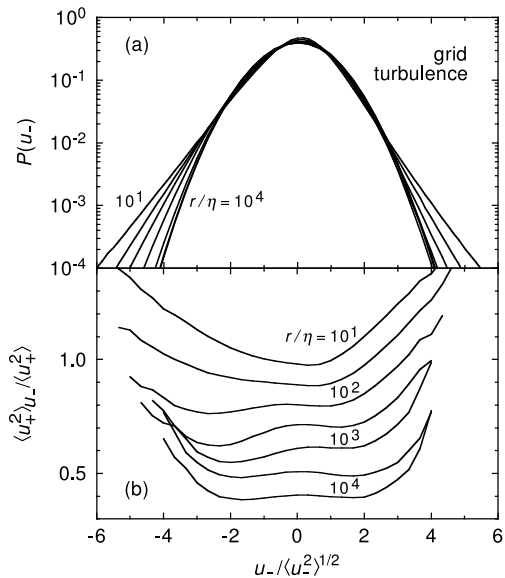}}
\caption{\label{f4} { Statistics for grid turbulence at $r/\eta = 10^1$, $3 \times 10^1$, $10^2$, $3 \times 10^2$, $10^3$, $3 \times 10^3$, and $10^4$ as a function of $u_-/ \langle u_-^2 \rangle^{1/2}$. (a) Probability density distribution of $u_-$. (b) Conditional average $\langle u_+^2 \rangle_{u_-}$ normalized by $\langle u_+^2 \rangle$. The value at each $r$ is shifted by $0.1$.}}
\end{figure} 
%_______________________________________________________________________________________
\begin{figure}[htbp]
\resizebox{8cm}{!}{\includegraphics*[4.cm,14.7cm][16.cm,26.3cm]{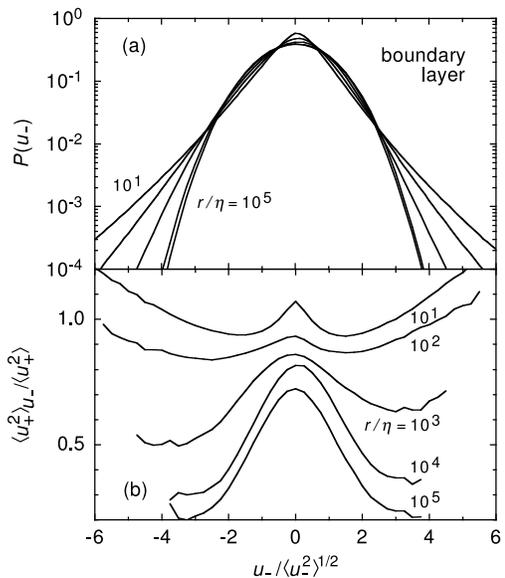}}
\caption{\label{f5} { Same as in Fig.~\ref{f4} but for boundary layer at $r/\eta = 10^1$, $10^2$, $10^3$, $10^4$, and $10^5$.}}
\end{figure} 
%_______________________________________________________________________________________
\begin{figure}[htbp]
\resizebox{8cm}{!}{\includegraphics*[4.cm,14.7cm][16.cm,26.3cm]{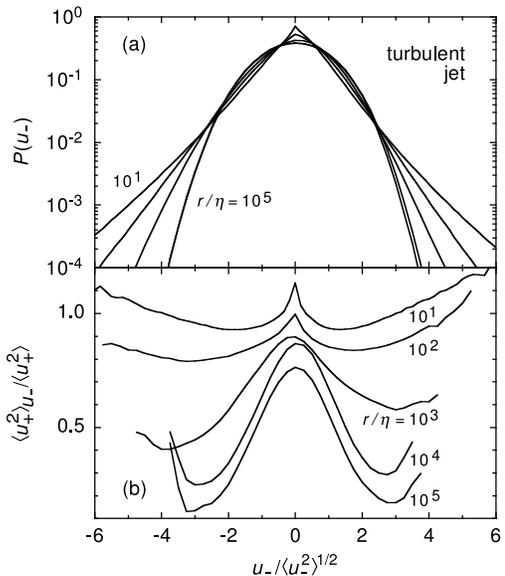}}
\caption{\label{f6} { Same as in Fig.~\ref{f4} but for jet at $r/\eta = 10^1$, $10^2$, $10^3$, $10^4$, and $10^5$.}}
\end{figure} 
%_______________________________________________________________________________________

\section{ENERGY TRANSFER RATE} \label{s5}

To understand how the mean rates of energy transfer $-5 \langle u_-^3(r) \rangle/4r$ and $15 \langle u_+^2(r) u_-(r) \rangle /r$ in Eqs.~(\ref{eq10}) and (\ref{eq11}) are determined as observed in Figs.~\ref{f1}(c), \ref{f2}(c), and \ref{f3}(c), we study fluctuations of $u_-(r)$ and $u_+^2(r) u_-(r)$.

Figure \ref{f4}(a) shows the probability density distribution of $u_-(r)$ in the grid turbulence. If $r \lesssim L_u \simeq 1 \times 10^3 \eta$, the distribution is skewed with a long tail in the negative side. This tail explains the observed positive value of $-5 \langle u_-^3(r) \rangle /4r$. The value vanishes if $r \gtrsim L_u$, where the distribution of $u_-(r)$ is symmetric.

Figure \ref{f4}(b) shows the average $\langle u_+^2(r) \rangle _{u_-}$ conditioned on $u_-(r)$ in the grid turbulence. If $r \lesssim L_u$, $\langle u_+^2(r) \rangle _{u_-}$ is greater at $u_- = u_{\ast} > 0$ than it is at $u_- = -u_{\ast} < 0$. This asymmetry overbalances the negative skewness of $u_-$ and explains the observed positive value of $15 \langle u_+^2(r) u_-(r) \rangle /r$. The value vanishes if $r \gtrsim L_u$, where $\langle u_+^2(r) \rangle _{u_-}$ is symmetric.

The same $r$ dependences are observed for the boundary layer in Fig.~\ref{f5} with $L_u \simeq 3 \times 10^3 \eta$ and for the jet in Fig.~\ref{f6} with $L_u \simeq 7 \times 10^3 \eta$.

Besides the asymmetry, $\langle u_+^2(r) \rangle _{u_-}$ has features that are not common to the grid turbulence, boundary layer, and jet [Figs.~\ref{f4}(b), \ref{f5}(b), and \ref{f6}(b)]. The configuration for turbulence production affects large-scale motions, which in turn affect $u_+^2(r)$ at any $r$. However, not affected is the mean rate of energy transfer $15 \langle u_+^2(r) u_-(r) \rangle /r$ because $u_+^2(r)$ correlates with $u_-(r)$ only through motions at the scale $r$ (Sec.~\ref{s3}).

The local rate of energy transfer is known to fluctuate markedly.\cite{dr90,ok92} This fluctuation could be related to those of $-5 u_-^3(r)/4r$ and $15 u_+^2(r) u_-(r)/r$, although they are not exactly identical to the local rate.

Phenomenologically, when an eddy typical for the size $r$ deforms and breaks apart into the smaller daughter eddies, the deformation energy $\langle u_-^2(r) \rangle$ of the mother eddy is transferred to those of the daughter eddies. Since the time scale is $r/\langle u_-^2(r) \rangle^{1/2}$, this offers a conceptual explanation for the mean rate of energy transfer $-5 \langle u_-^3(r) \rangle /4r$. The mother eddy also has energy $\langle u_+^2(r) \rangle$ for the advection, which is also transferred to those of the daughter eddies. With the time scale $r/\langle u_-^2(r) \rangle^{1/2}$, this explains the mean rate in terms of $15 \langle u_+^2(r) u_-(r) \rangle /r$. These two explanations are equivalent if we consider eddies with all the sizes, since the advection $u_+$ of an eddy reflects the deformations $u_-$ of the surrounding larger eddies. In fact, we have $-5 \langle u_-^3(r) \rangle /4r = 15 \langle u_+^2(r) u_-(r) \rangle /r$.

\section{SCALE DEPENDENCE OF MOMENTS} \label{s6}

The $r$ dependences of $\langle u_+^n(r) \rangle$ and $\langle v_+^n(r) \rangle$ are distinct from those of $\langle u_-^n(r) \rangle$ and $\langle v_-^n(r) \rangle$. For example, while $\langle u_-^2(r) \rangle$ varies by several orders, $\langle u_+^2(r) \rangle$ varies only from $\langle u^2 \rangle /2$ to $\langle u^2 \rangle$ [Figs.~\ref{f1}(a), \ref{f2}(a), and \ref{f3}(a)].\cite{ty07} This range is too narrow to exhibit a power-law scaling $\propto r^{\zeta}$ such as those known for $\langle u_-^n(r) \rangle$.\cite{k41a,sa97}

The $r$ dependences of $\langle u_+^n(r) \rangle$ and $\langle v_+^n(r) \rangle$ are also subject to large-scale motions that depend on the configuration for turbulence production. For example, although exactly isotropic turbulence exhibits $\langle u_+^3(r) \rangle = 0$ at all $r$, our grid turbulence exhibits some variation of $\langle u_+^3(r) \rangle$. There still exist features that do not depend on the flow configuration, as demonstrated below by using long experimental data.

Figure \ref{f1}(b) shows the flatness factors in the grid turbulence, e.g., $\langle u_+^4(r) \rangle / \langle u_+^2(r) \rangle^2$. With a decrease in $r$, those of $u_-(r)$ and $v_-(r)$ increase as a result of the small-scale intermittency.\cite{sa97} Those of $u_+(r)$ and $v_+(r)$ are constant within $\pm 1$\%. The reason is the predominance of large-scale motions. In fact,
%_______________________________________________________________________________________
\begin{equation}
\label{eq12}
{ \everymath{\displaystyle}
  \frac{\langle u_+^4(r) \rangle}{\langle u_+^2(r) \rangle^2} \rightarrow 
  \left\{ \begin{array}{ll}
             \frac{\langle u^4 \rangle}{\langle u^2 \rangle ^2}              &\quad \mbox{as} \quad r \rightarrow 0, \\
             \rule{0cm}{5ex}
             \frac{\langle u^4 \rangle}{2\langle u^2 \rangle ^2}+\frac{3}{2} &\quad \mbox{as} \quad r \rightarrow \infty .
          \end{array} \right.
}
\end{equation}
%_______________________________________________________________________________________
These asymptotic values are the same if $\langle u^4 \rangle / \langle u^2 \rangle ^2 =3$, which is close to the case of the grid turbulence (Table \ref{t1}). However, in more detail, the flatness factors of $u_+(r)$ and $v_+(r)$ vary with the scale $r$, exhibiting maxima at $r \simeq L_u$ and minima at $r \simeq 10^{-1} L_u$. Albeit of small amplitude, these variations are still above the statistical uncertainties that are seen as rapid fluctuations at $r \gtrsim 10^1 L_u$. The same variation exists for the flatness factor of $v_+(r)$ in the boundary layer and jet  [Figs.~\ref{f2}(b) and \ref{f3}(b)].\cite{note6}

For the observed variations of the flatness factors, we attempt an explanation, which is not complete but would serve as a guide for future studies. When $r$ is not too large, $u_+(r)$ represents the advection of an eddy with size $r$, which is due to sweeping by nearby eddies. If each of them has the same sweeping direction, the corresponding $u_+$ value tends to lie at the tail of the $u_+$ distribution. Since eddies with sizes $\lesssim r$ do not contribute to the sweeping, an increase in $r$ shortens the above tail. The flatness factor becomes small. With a further increase in $r$, eddies with sizes $\gtrsim r$ become sparse. The $u_+$ distribution becomes enhanced at $u_+ \simeq 0$, i.e., cases of no eddies. The flatness factor becomes large. Finally, when $r$ exceeds the typical size $L_u$ of energy-containing eddies, their random relative motions become dominant. The flatness factor tends toward the Gaussian value of 3 [Eq.~(\ref{eq12})].

\section{CONCLUDING REMARKS} \label{s7}

The velocity averages $u_+(r)$ and $v_+(r)$ represent motions over scales $\ge r$, as long as $r$ is not too large for motions at scales $< r$ to be negligible, i.e., for the velocity differences $u_-(r)$ and $v_-(r)$ to be safely assumed as representatives of motions at the scale $r$. Phenomenologically, while $u_-(r)$ and $v_-(r)$ represent the deformation and rotation of an eddy with size $r$, $u_+(r)$ and $v_+(r)$ represent the advection of such an eddy.

The motions over scales $\ge r$ are defined by $\bar{u}(r)$, i.e., velocity smoothed over the scale $r$ [Eq.~(\ref{eq5})]. When $r$ is not too large, we observe $\langle u_+^2(r) \rangle \simeq \langle \bar{u}^2(r) \rangle$ [Figs.~\ref{f1}(a), \ref{f2}(a), and \ref{f3}(a)].

We have found an exact scale-by-scale energy budget equation for $u_+^2$ [Eq.~(\ref{eq10b}) or (\ref{eq11b})], which involves the relation found by Hosokawa\cite{ho07} [Eq.~(\ref{eq3b})] as the mean rate of energy transfer across the scale $r$. This equation describes the same energy budget as the equation for $u_-^2$ [Eq.~(\ref{eq10a}) or (\ref{eq11a})]. Between them, the corresponding terms have exactly the same values.

We have also found that the flatness factors of $u_+(r)$ and $v_+(r)$ vary with $r$ in a manner that is independent of the configuration for turbulence production [Figs.~\ref{f1}(b), \ref{f2}(b), and \ref{f3}(b)]. Thus, albeit subject to large-scale motions that depend on the flow configuration, $u_+$ and $v_+$ have universal features.

The analogues of $u_+$ and $u_-$ exist in the orthonormal wavelet transformation, where $u(x)$ is transformed as\cite{yo91,m91,m99}
%_______________________________________________________________________________________
\begin{subequations}
\begin{equation}
\label{eq13}
\sum_{m=-\infty}^{+\infty} \tilde{u}_{S,N,m}S_{N,m}(x) + \sum_{n=N}^{+\infty} \, \sum_{m=-\infty}^{+\infty} \tilde{u}_{W,n,m}W_{n,m}(x),
\end{equation}
%_______________________________________________________________________________________
with two self-similar families of basis functions
%_______________________________________________________________________________________
\begin{align}
& S_{n,m}(x) = 2^{n/2} S(2^n x-m), \\
& W_{n,m}(x) = 2^{n/2} W(2^n x-m). 
\end{align}
\end{subequations}
%_______________________________________________________________________________________
They are localized around $x = 2^{-n}m r_{\ast}$, where $r_{\ast}$ is some unit scale. While $S_{n,m}(x)$ represents motions over scales $r \gtrsim 2^{-n} r_{\ast}$, $W_{n,m}(x)$ represents motions at the scale $r \simeq 2^{-n-1} r_{\ast}$. If $2^{-n} r_{\ast}$ is not too large, the transform $\tilde{u}_W$ is analogous to the velocity difference $u_-$.\cite{yo91,m99} The transform $\tilde{u}_S$ is analogous to the velocity average $u_+$. There exist various wavelets based on various definitions of the scales. Those for Haar's wavelet\cite{m91,m99,h10} are the same as defined for the present study [Eq.~(\ref{eq5})].

Tatsumi and Yoshimura\cite{ty04,ty07} proposed a closure approximation that $u_+$ is statistically independent of $u_-$ at each scale, which is interpreted as an approximation of statistical independence between the advection $u_+(r)$ and deformation $u_-(r)$ of eddies with each size $r$. This is exact in many cases [Eq.~(\ref{eq7b})]. An exception is the correlation of $u_+^2$ with $u_-$ studied here [Eq.~(\ref{eq2})].\cite{ho07}

There is a controversy as to whether or not small-scale motions in the dissipative and inertial ranges are statistically independent of large-scale motions in the energy-containing range.\cite{sd98,ho07,kt08,m06,ll59,o62,p93} To address this issue, $u_-^n(r)$ at small $r$ is often compared with $u_+^m(r)$ or $u^m$, by assuming that they represent the large-scale motions. Being actually a representative of motions over scales $\ge r$, $u_+^m(r)$ could correlate with $u_-^n(r)$ through motions at the scale $r$ [Eq.~(\ref{eq2})]. The situation is the same for $u^m$, which corresponds to $u_+^m(r)$ for $r \rightarrow 0$ and hence could correlate with $u_-^n(r)$. Such correlations are spurious as those between small- and large-scale motions. In particular, the large-scale motions are not responsible to $\langle u_+^2(r) u_-(r) \rangle \ne 0$ for $r$ in the inertial range [Eq.~(\ref{eq3b})].

To conclude this paper, we underline that further studies of $u_+$ and $v_+$ are promising.\cite{sd98,ty04,ty07,ho07,kt08,hi02,ge07} For the above and other issues, $u_+$ and $v_+$ could reveal new features and offer new insights. This is because the description of turbulence by $u_+$ and $v_+$ differs from the existing description by $u_-$ and $v_-$, and also because $u_+$ and $v_+$ satisfy exact equations like Eq.~(\ref{eq10b}) that serve as the firm theoretical basis.

\begin{acknowledgments}
We are grateful to K. Hashimoto and Y. Kawashima for continued collaborations as well as to I. Hosokawa, T. Gotoh, M. Takaoka, T. Tatsumi, S. Toh, and A. Tsinober for useful comments. This research was supported in part by KAKENHI (C) 22540402.
\end{acknowledgments}

%\clearpage


\begin{thebibliography}{999}
\bibitem{note0} Turbulence in this hypothetical space $x$ is still realistic. Local values of $u(x)$ and $v(x)$ represent local motions of the actual turbulence. They are continuously connected up to largest scales (Ref. \onlinecite{m06}).
\bibitem{note2} There could exist turbulence that is not homogeneous but is still isotropic around some specific position $x_{\ast}$. By setting $x_{\ast}$ to be the midpoint of $x$ and $x+r$ and by taking ensemble averages, we obtain the same relations as in Eqs.~(\ref{eq2}) and (\ref{eq7}), although they do not necessarily hold around the other positions.
\bibitem{note3} For this isotropic turbulence, we also obtain $\partial_r [r^2 \langle u_+^2(r) \rangle ] = 2r \langle v_+^2(r) \rangle$ from $\partial_r [r^2 \langle  u_-^2(r) \rangle ] = 2r \langle v_-^2(r) \rangle$, $\langle u^2 \rangle = \langle v^2 \rangle$, and Eq. (\ref{eq7a}).
\bibitem{note4} Normalized by different estimates of $\langle \varepsilon \rangle$ to compensate for unavoidable artifacts in our experiments. The measurement of the $u$ component was not so reliable at smallest scales (Ref. \onlinecite{mht09}). The grid turbulence was not so isotropic at largest scales. Then, the most reliable estimate of $\langle \varepsilon \rangle$ is $15 \nu \langle (\partial _x v)^2 \rangle /2$ in Table \ref{t1}.
\bibitem{note5} The right-hand sides of Eqs.~(\ref{eq10a}) and (\ref{eq10b}) are determined at each scale $r$. Although $u_+^2(r)$ represents motions over scales $\ge r$, any information about scales $> r$ has been lost from $\partial_r \langle u_+^2(r) \rangle$, i.e., differential with respect to $r$. Such information has been also lost from $\langle u_+^2(r)u_-(r) \rangle$ because $u_+^2(r)$ correlates with $u_-(r)$ only through motions at the scale $r$ (Sec.~\ref{s3}).
\bibitem{note6} For these boundary layer and jet, $\langle u_+^4(r) \rangle / \langle u_+^2(r) \rangle^2$ is not shown. Its values for $r \rightarrow 0$ and $r \rightarrow \infty$ are too different because $\langle u^4 \rangle /\langle u^2 \rangle^2$ is too different from $3$ [Table \ref{t1}, see also Eq.~(\ref{eq12})], although there appears to exist the same variation as observed for the grid turbulence.
\end{thebibliography}
\end{document}